\documentclass[aps,prl,preprintnumbers,groupedaddress,reprint]{revtex4-1}
\usepackage{blindtext}
\usepackage{graphicx}
\usepackage{slashed}
\usepackage{subcaption}
\usepackage[export]{adjustbox}
\usepackage{todonotes} 
\usepackage{amsmath}
\usepackage[normalem]{ulem}

\newcommand{\nn}{\nonumber}
\newcommand{\bea}{\begin{eqnarray}}
\newcommand{\eea}{\end{eqnarray}}

\begin{document}

\preprint{\vbox{
		\hbox{MIT-CTP 5562}
}}

\title{Power Counting to Saturation}
\author{Iain W.~Stewart$^{1}$ and Varun Vaidya$^{2}$ }
\affiliation{${}^1$ Center for Theoretical Physics, Massachusetts Institute of Technology, Cambridge, MA 02139, USA}
\affiliation{${}^2$ Department of Physics, University of South Dakota, Vermillion, SD, 57069}

\begin{abstract}

We present a description of saturation in small $x$ deep inelastic scattering from power counting in a top-down effective theory derived from QCD.  
A factorization formula isolates the universal physics of the nucleus at leading power in $x$. The onset of saturation is then understood as a breakdown in the expansion in an emergent power counting parameter, which is 
defined by the matrix element of a gauge invariant operator.  We identify a new radiation mode, which enables us to extend previous literature by distinguishing the appearance of the saturation scale from the transition to non-linear evolution. 

\end{abstract}

\maketitle

One of the primary objectives of a future Electron Ion Collider is an investigation of the partonic structure of hadrons or large nuclei at high energies.  The partonic density at such high energies is enhanced due to quantum radiative corrections leading to the phenomenon called saturation \cite{Gribov:1983ivg, Mueller:1989st,Mueller:1985wy}. The current understanding of saturation is via the Color Glass Condensate (CGC) \cite{Iancu:2003xm, Gelis:2010nm,Kovchegov:2012mbw} which treats the hadron/nucleus as a classical source for small $x$ gluons. 
This is often thought of as a bottom-up effective theory \cite{McLerran:1998nk}, whose degrees of freedom consist of color sources $\rho$ at large $x$ with momenta above a cutoff, $k^+ > \Lambda^+$, and gauge fields $A^\mu$ at small $x$ with $k^+ < \Lambda^+$. 
The input is a gauge invariant distribution $W_{\Lambda^+}[\rho]$ which determines the probability of a configuration $\rho$, 
and at lowest order one solves classical Yang-Mills equations for $A^\mu$ with $\rho$ as a source. 
This formalism incorporates 
soft gluon radiative corrections via the non-linear Balitsky-JIMWLK or the Balitsky-Kovchegov (BK)(at large $N_C$) \cite{Balitsky:1995ub,Jalilian-Marian:1997qno,Jalilian-Marian:1997jhx,Kovner:2000pt,Iancu:2000hn,Iancu:2001ad,Ferreiro:2001qy,Kovchegov:1999yj} equations which describe the evolution in the cutoff $\Lambda^+$. 

Our understanding of deep inelastic scattering (DIS) at large $x$ rests on the notion of non-perturbative factorization which enables us to define operators which separate the universal physics of the hadron from process dependent short distance physics~\cite{Sterman:1986aj,Catani:1989ne}. This can be thought of as a top-down effective field theory derived from QCD~\cite{Bauer:2002nz}, where the power counting expansion parameter is $\Lambda_{\rm QCD}/Q$.  Here $Q^2= -q^2 >0$ where $q^\mu$ is the momentum transfer of the virtual photon in the electron-proton collision, $x = Q^2/(2 P\cdot q)$ where $P^\mu$ is the initial proton's momentum, and we define $H = (P+q)^2$, the photon-nucleus invariant mass.

The goal of this letter is to derive an effective field theory description of small $x$ saturation physics from the top down, and connect it to the CGC formalism. 
Starting from QCD we setup a factorization framework for DIS at small $x$ at all orders in the strong coupling and leading power in $x$.  This leads to a definition for the universal physics of the  nuclear medium as an operator matrix element in the nuclear state. Next we show that we can understand  the appearance of the saturation scale in terms of a breakdown in an emergent power counting  parameter that appears due to multiple probe-medium interactions. 
Finally we give a {\it distinct} power counting argument to understand how the form of the factorization changes as we enter the regime with non-linear evolution in $\ln x$, as we approach the Froissart bound.
Apart from the kinematical scales that we will discuss in the next section, we foreshadow the importance of three emergent length scales that are crucial for this description. The first is $L_D$, the average distance beyond which color sources in the medium are uncorrelated. The second is the time scale $t_c$ over which the partons in the probe maintain quantum coherence. Finally we have the mean free path of the dipole probe $\lambda_{\text{mfp}}$ which is the average distance between successive interactions of the probe with the medium.  
\paragraph{{\bf Kinematics and Factorization:}}

We want to describe the small $x \ll 1$  region or equivalently $\Lambda_{QCD}^2 \ll Q^2 \ll 2P\cdot q$, which is the high energy limit of QCD where $H \approx 2P\cdot q$.
Without loss of generality, we can orient our axes so that $q^{\mu}$ has no transverse momentum and work in a frame with
$(q^-, q^+ ,  q^{\perp}) = ( Q,  -Q, 0^{\perp}) $,
where we decompose
$q^{\mu} = q^- \frac{n^{\mu}}{2}+q^+\frac{\bar{n}^{\mu}}{2}+q^{\perp}$, with $n^{\mu} \equiv (1, 0, 0 ,1)$, 
$ \bar{n}^{\mu} \equiv(1,0,0,-1) $, and $\perp$ refers to momenta in the $x$-$y$ plane.
For a proton of mass $M_p$, the momentum components in this frame are $(M_p^2Q/H,  H/Q,  0)$.
 
The dominant interaction is the photon fluctuating into a quark-anti quark dipole which then scatters off the hadron/nucleus~\cite{Gribov:1968gs}.  The medium partons exchange a transverse momentum of order Q with  the dipole and scatter by an angle $\theta \sim q_T/E \sim  Q^2/H = x \ll 1$ which is the power counting parameter of the Effective Field Theory(EFT). 

In terms of this expansion parameter, the partons in the dipole scale as the soft mode
$p_s \sim  \frac{H}{Q}\left(x, x, x\right)$
while the partons in the medium scale as an $\bar n$ collinear mode
$p_{\bar n} \sim  \frac{H}{Q}\left(x^2, 1, x\right)$.
Both these modes are propagating degrees of freedom with an invariant mass $p^2 \sim Q^2$ . This stage of the EFT is therefore designed to describe IR physics at the scale Q while separating out the physics at the scale $H/Q$.
The small angle scattering between these modes is mediated by a Glauber mode whose scaling is fixed by the constraint that both the dipole and medium partons must retain their scaling in any interaction. Therefore, we can write 
$p_G \sim  \frac{H}{Q}\left(x^2,x,x\right) $.
These same EFT modes are used in~\cite{Neill:2023jcd} to derive small-$x$ factorization for DIS with applications to $\ln x$ resummation for DIS coefficient functions and anomalous dimensions. 

\begin{figure}
\centering
\includegraphics[width=0.7\linewidth]{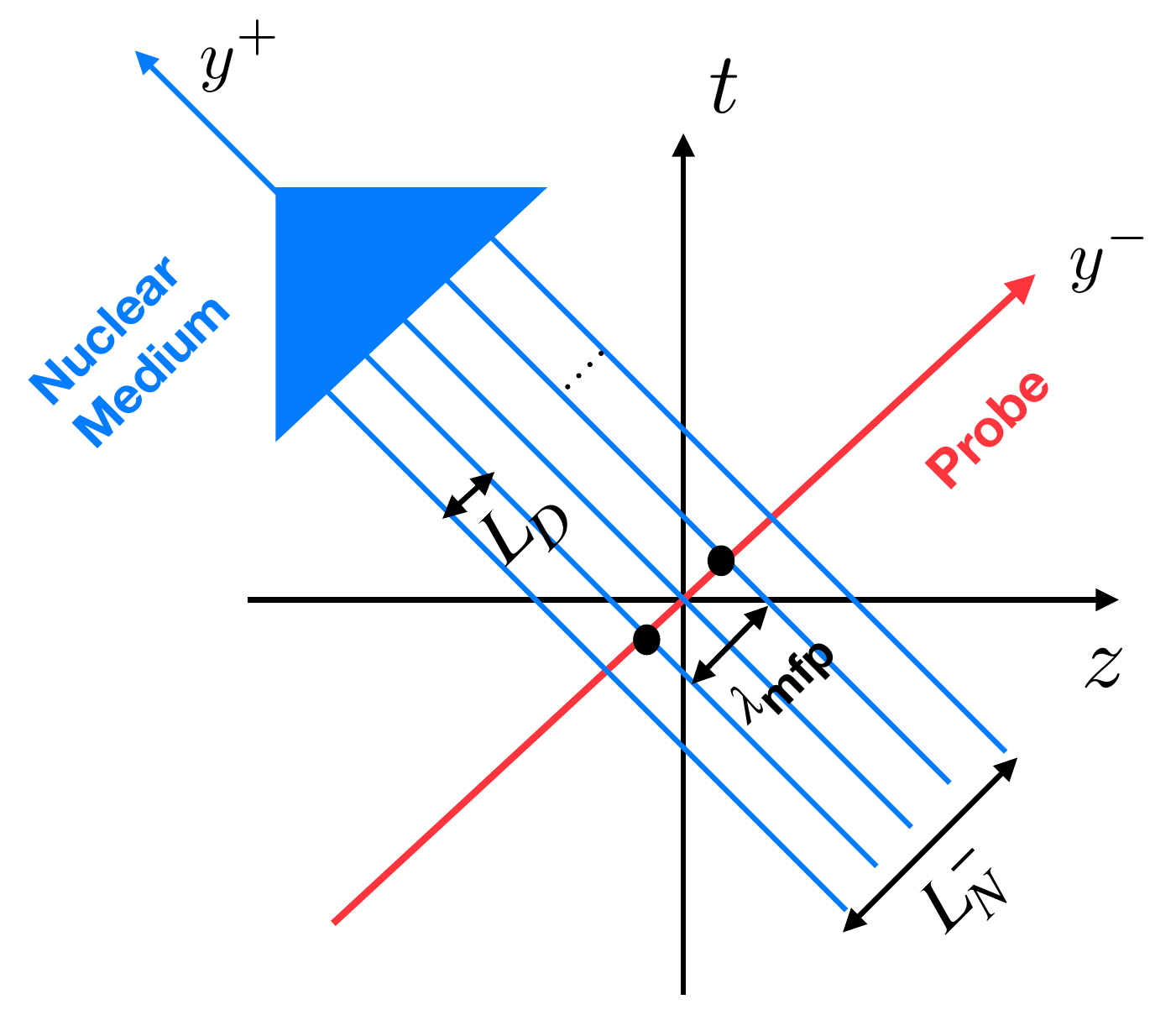}
\caption{Interaction of the probe with the medium.}
\label{Smallx}
\end{figure}

The QCD action can be systematically expanded in $\theta  =x$ within the Soft Collinear Effective Theory (SCET)~\cite{Bauer:2002aj,Bauer:2003mga,Bauer:2000yr,Bauer:2001ct,Bauer:2002nz}, as shown in~\cite{Rothstein:2016bsq}. This allows us to write an effective Hamiltonian at leading power in $x$
\bea
H_{\text{eff}} = H_s+H_{\bar n}+ H_G^{\bar n s}+ e_q \bar q_s\gamma_{\mu} q_s L^{\mu}+ \ldots
\label{eq:Heff}
\eea 
This has an effective hard interaction mediated by a photon between the electron current $L^{\mu}$ and the soft quark current, where $e_q$ is the electric charge of the quark. 
Soft and collinear modes of SCET are decoupled in the 
soft $H_s$ and collinear $H_{\bar n}$ Hamiltonians, but couple through the Glauber Hamiltonian  density$\mathcal{H}_G^{\bar n s}$ that mediates forward sacttering and has the form
\begin{align}
\label{EFTOp}
\mathcal{H}_G^{\bar n s} 
 &= C_G\!\! \sum_{i,j \in \{q,g\}} \mathcal{O}_{\bar ns}^{ij}
 \,, 
 & \mathcal{O}_{\bar ns}^{ij} & =\mathcal{O}_{\bar n}^{iB}\frac{1}{\mathcal{P}_{\perp}^2}\mathcal{O}_s^{jB} 
 \,,
\end{align}
where $C_G(\mu)=8\pi\alpha_s(\mu)$ and ${\cal P}_\perp^\mu$ is a derivative operator in the $\perp$ direction. 
The gauge invariant $\bar n$ collinear and soft operators are
\small
\bea
\mathcal{O}_{\bar n}^{qB} = \bar{\chi}_{\bar n}T^B\frac{\slashed{n}}{2}\chi_{\bar n}, 
 \  \   \    
\mathcal{O}_{\bar n}^{gB}=  \frac{i}{2}f^{BCD}\mathcal{B}_{\bar n \perp\mu}^C\frac{n}{2}\cdot(\mathcal{P}+\mathcal{P}^{\dagger})\mathcal{B}_{\bar n \perp}^{D\mu}
\nn\\
\mathcal{O}_{s}^{qB} = \bar{\chi}_{s}T^B\frac{\slashed{\bar{n}}}{2}\chi_{s}, 
 \   \   \    
\mathcal{O}_{s}^{gB}=  \frac{i}{2}f^{BCD}\mathcal{B}_{s \perp\mu}^C\frac{\bar n}{2}\cdot(\mathcal{P}+\mathcal{P}^{\dagger})\mathcal{B}_{s \perp}^{D\mu} 
\,,
\nn\\
  \label{eq:Opbarn}
 \eea
 \normalsize
which are built out of the gauge invariant building blocks
\small
\begin{align}
 & \chi_{\bar n} = W_{\bar n}^{\dagger}\xi_{\bar n}, 
  \   \   \   \    
  W_{\bar n} = \text{FT} \  {\bf P} \exp \Big\{ig\int_{-\infty}^0 ds n\cdot A_{\bar n}(x+ns)\Big\}  ,
  \nn\\
 & \chi_{s} = S_{\bar n}^{\dagger}\xi_{s},  
   \   \   \   \    
   S_{\bar n} = \text{FT} \ {\bf P} \exp \Big\{ig\int_{-\infty}^0 ds\bar n\cdot A_{S}(x+s\bar{n} )\Big\}
   , \nn\\
 & \mathcal{B}_{\bar n \perp}^{C\mu} T^C = \frac{1}{g}\Big[W_{\bar n}^{\dagger}iD_{\bar n \perp}^{\mu}W_{\bar n}\Big],    \  \  
 \mathcal{B}_{s\perp}^{C\mu} T^C = \frac{1}{g}\Big[S_{\bar n}^{\dagger}iD_{S \perp}^{\mu} S_{\bar n}\Big] 
 .
\end{align}
\normalsize
FT stands for Fourier transform. 
These operators encode bare quark and gluons dressed by Wilson lines.
Since the dipole probe can interact multiple times with the medium, we can think of it as an open quantum system evolving in a nuclear environment.  We therefore follow the evolution of the reduced density matrix for our probe. 
The initial state density matrix can be written as
\bea 
\rho(0)= |e^-\rangle  \langle e^-| \otimes \rho_A 
\eea
where the electron is disentangled from the nuclear medium ($\rho_A$). We will evolve this with $H_{\text{eff}}$ and measure the properties of the electron,  being inclusive over the hadronic final states. In DIS we write the cross section differential in $Q^2$ and $x$, which we indicate with a measurement $\mathcal{M}$ which acts on the phase space of the final state $e^-$.  Since the Glauber mode couples the soft and collinear sectors, we have to expand out the density matrix element order by order in the Glauber Hamiltonian, $H_{\bar n s}^G$ in order to carry out soft-collinear factorization,
\bea
\Sigma &\equiv & \lim_{t\rightarrow \infty}\text{Tr}[e^{-iH_{\text{eff}}t}\rho(0)e^{iH_{\text{eff}}t}\mathcal{M}] = \sum_{i}\Sigma^{(i)}
\label{eq:Opacity}
\eea 
This is an expansion in the number of interactions of the probe with the medium and is often referred to as the opacity expansion~\cite{Gyulassy:2000fs}.  
Taking the confinement length scale $\Lambda_{\rm QCD}^{-1}$ and boosting to obtain a scale relevant for the size of a nucleon in the $y^-$ direction, $L_{D}~\sim M_p Q/(H\Lambda_{\rm QCD})$.  In reality $L_D$ might be substantially larger due to inter-nucleon interactions, especially in larger nuclei. We also consider the mean free path of the probe through the medium, $\lambda_{\text{mfp}}\sim xQ/(\alpha_s(Q)^2\Gamma)$, which determines how often the probe interacts with the medium, and depends on the interaction strength and medium density. Here $\Gamma$ is the nucleon density per unit area, where $\Gamma \sim \Lambda_{\rm QCD}^2$ for nucleons stacked next to one another.
We work in the regime $\lambda_2 = L_D/\lambda_{\text{mfp}}\sim \Gamma/Q^2 \ll 1$ where successive interactions of the probe with the medium happen with color uncorrelated partons. 
The assumption $\lambda_2\ll 1$ dictates that only even powers of $H^G_{\bar n s}$ contribute so as to have the same number of Glauber exchanges on both sides of the cut.  $\lambda_2 \ll 1$ implies that each  successive Glauber exchange  happens with an  independent scattering center displaced in the longitudinal direction by $\lambda_{mfp} \gg L_D$.  Therefore at the amplitude squared level , each Glauber exchange is accompanied by its complex conjugate, and odd number of Glauber exchanges are excluded .
The factorization formula at $O(2m)$ in this series is
\begin{widetext}
\begin{eqnarray}
\Sigma^{(2m)}&=&  \frac{|C_G(\mu)|^{2m}}{Q^4} \Bigg[\int d^+p_e \mathcal{M}\Bigg] L_{\alpha \beta} \int_{\bar y \in \text{Nucleus}} d\bar{y}^+ d^2\bar{y}_{\perp}\nn\\
&&\text{Im}\Bigg\{ \Bigg[\Pi_{i=1}^{m}\int_0^{L_N^-} d\bar{y}_i^-\Theta(\bar{y}_{i}^--\bar{y}_{i+1}^-)\int \frac{d^2 k_{i}^{\perp}}{(2\pi)^2} \mathcal{B}_{\bar n} (k_{i}^{\perp},\bar{y}_i^-,\bar{y}^+,\bar{y}^{\perp};\nu)\Bigg] S_{m}^{\alpha \beta}(k_{1}^{\perp}, ... k_{m}^{\perp};\bar{y}_1^-,...\bar{y}_m^-;\mu, \nu)\Bigg\}
 .
\label{eq:NLOG}
\end{eqnarray}
 \end{widetext}
It is written in terms of $m$ copies of an $\bar n$-collinear medium structure function $\mathcal{B}_{\bar n}$ and a soft function $S_m^{\alpha\beta}$ describing the evolution of
the Dipole probe.
The $\lambda_2\ll 1$ expansion leads to the product of ${\cal B}_{\bar n}$ functions, rather than a single multi-interaction $\bar n$ correlation function.
$\int d^+p_e$ is the phase space integral over the final state electron and $L_{\alpha \beta}$ is the leptonic tensor.  The coordinate $\bar y$ indicates the position inside the nucleus. The physical picture is that the successive interactions of the probe dipole occur with color uncorrelated partons in the nucleus at a single value of $\bar{y}^+,\bar{y}^{\perp}$but at increasing values of $\bar y^-$ as shown in Fig.\ref{Smallx}.  The nucleus extends over a length $L_N^-$ in the $y^-$ direction.  $k^{\perp}_i$ is the transverse momentum exchanged in the $i^{th}$ interaction.  
The separation of collinear-soft physics leads to rapidity divergences, for which we use the dimensional regularization like rapidity regulator of~\cite{Chiu:2012ir} with the corresponding renormalization scale $\nu$.
$\mathcal{B}_{\bar n}$ is a function of $\bar y$ so that it can also describe an inhomogeneous medium. It is given by a current-current correlator in the nuclear background
 \bea
&& \mathcal{B}_{\bar n}(k_{\perp},\bar{y};\nu) = \frac{1}{2}\frac{1}{k_{\perp}^2} (N_C^2-1)\int d^2\hat{y}_{\perp}d\hat{y}^- e^{i k_{\perp}\cdot  \hat {y}^{\perp}}\nn\\
 &\times &\text{Tr}\Big[ O^A_{\bar n}\left(\frac{1}{2}(-\hat{y}+\bar{y})\right)O^A_{\bar n}\left(\frac{1}{2}(\hat{y}+\bar{y})\right)\rho_A \Big]
  .
 \label{eq:MediumF}
\eea
This captures the $universal$ physics of the medium to all orders in perturbation theory and is independent of any measurements that are made on the probe. $\mathcal{O}_{\bar n}^A$ can be thought of as the operator analog of the color source function $\rho$ in the CGC approach \cite{McLerran:1998nk}. The medium structure function does not have any UV divergences but obeys a renormalization group equation in rapidity which is the Balitsky–Fadin– Kuraev–Lipatov (BFKL) \cite{Kuraev:1977fs,Balitsky:1978ic} equation at one loop.
\small
\begin{align}
 &\frac{d \mathcal{B}_{\bar n}(k_{\perp},\bar y;\nu)}{d\ln \nu} = - \int d^2u_{\perp}K_{\text{BFKL}}(\vec{k}_{\perp},\vec{u}_{\perp})\mathcal{B}_{\bar n}(u_{\perp},\bar{y};\nu)\nn\\
 &= -\frac{\alpha_s N_C}{\pi^2} \int d^2u_{\perp} \Bigg[\frac{\mathcal{B}_{\bar n}(u_{\perp},\bar{y};\nu)}{(\vec{u}_{\perp}-\vec{k}_{\perp})^2} -\frac{k_{\perp}^2\mathcal{B}_{\bar n}(k_{\perp},\bar{y};\nu) }{2u_{\perp}^2(\vec{u}_{\perp}-\vec{k}_{\perp})^2}\Bigg]
  .
\end{align}
\normalsize
Its natural rapidity scale corresponds to $\nu\sim H/Q$ for which logarithms in this function are minimized.
By RG consistency, we have
\small
\begin{align}
 &\frac{d S^{\alpha\beta}_{m}(k_{1\perp},...k_{m\perp},\{\bar{y}_i^-\};\mu,\nu)}{d\ln \nu}= \sum_{i=1}^n \int d^2u_{\perp}K_{\text{BFKL}}(u_{\perp},k_{i\perp}) \nn\\
 &\times S^{\alpha \beta}_{m}(k_{1\perp},..k_{i-1\perp},u_{\perp}, k_{i+1\perp},..k_{m\perp},\{\bar{y}_i^-\};\mu,\nu)
.
\end{align}
\normalsize
The natural rapidity scale for this function is $\nu \sim Q$.
We can use the rapidity RG equation to resum the large logarithms in $x$ by solving the BFKL equation, running the function $\mathcal{B}_{\bar n}$ from $\nu \simeq s/Q$ to $\nu \simeq Q$ while setting $\nu \simeq Q $ in the Dipole function, leading to an RG improved function $\mathcal{B}_{\bar n}^{\text{R}}(k_\perp,\bar y,x)$.
At leading log, we can resum the series in Eq.(\ref{eq:Opacity}) to all orders in the Glauber interactions
\small
\begin{align}
 \Sigma= \int\!\! d^2\bar{y}^{\perp}d\bar{y}^+\!\! \int\!\! d^2b\: \mathcal{F}(\vec{b},Q)\Bigg\{1-{\bf P}e^{-\mbox{$\int_0^{L^-_N}$} \frac{d\bar{y}^-}{\lambda_{\text{mfp}}\left(\vec{b},Q,\bar{y}\right)}}\Bigg\}
 ,
\label{eq:GGM}
\end{align}
\normalsize
which is a path ordered exponential with $\lambda_{\text{mfp}}$ given by a RG improved medium structure function $\mathcal{B}_{\bar n}^{\text{R}}$,
\small
\begin{align} \label{eq:lammfp}
\lambda^{-1}_{\text{mfp}}(\vec{b},Q,\bar{y}) &= \!\! \int\!\!\! \frac{d^2k_{\perp}}{(2\pi)^2k_{\perp}^2}|C_G(k_{\perp})|^2\mathcal{B}_{\bar n}^{\text{R}}(\vec{k}_{\perp},\bar y, x) 
 \nn\\
 &\qquad \times \Big[e^{i\vec{b} \cdot \vec{k}_{\perp}}-1\Big]
 .
\end{align}
\normalsize
Here $\lambda_{\text{mfp}}$ is an emergent length scale,  which can be interpreted as the mean free path of the probe in the medium in the $y^-$ direction.  
It depends on the local properties of the medium at $\bar y$.  The function $ \mathcal{F}(\vec{b},Q)$ includes the transverse component of the dipole function $S_n^{\alpha\beta}$ at tree level,
\small
\bea
 &&\mathcal{F}(\vec{b},Q)= 	\frac{1}{Q^3}\Bigg[\int d^+p_e \mathcal{M}\Bigg] L_{\alpha \beta}g_{\perp}^{\alpha \beta}\int d^2p_{\perp} d^2\bar{p}_{\perp}e^{i \vec{b} \cdot( \vec{p}_{\perp}+ \vec{\bar p}_{\perp})}\nn\\
 &&\int_0^1\!\! dz \left(z^2\!+\!(1\!-\!z)^2\right) \Bigg[\frac{\vec{p}_{\perp}}{Q^2z(1\!-\!z)\!
 +\vec{p}_{\perp}^2}+\frac{\vec{\bar p}_{\perp}}{Q^2z(1\!-\!z)\!+\vec{\bar p}_{\perp}^2}\Bigg]^2
 \label{eq:Wave}
 \nn
\eea
\normalsize
with the transverse size of the dipole $\vec{b}\sim 1/Q$,  as this is the only scale in this function.
The emergence of the mean free path automatically defines a new power counting parameter in the EFT 
\begin{align} \label{eq:lam1}
\lambda_1 = \int_0^{L^-_N} \frac{d\bar{y}^-}{\lambda_{\text{mfp}}\left(\vec{b},Q,\bar{y}\right)}
  \,.
\end{align}

The appearance of the saturation scale is just the breakdown of the power expansion in this parameter, when $\lambda_1\sim 1$. Given the fact that the transverse size of the dipole $\vec{b}$ is conjugate to the scale Q, we can define the saturation momentum scale $Q_s(\bar y)$ via
\bea
 \int_0^{L^-_N} \frac{d\bar{y}^-}{\lambda_{\text{mfp}}\left(
 |\vec{b}|= 1/Q_s,
 Q_s,\bar{y}\right)} =1 
 \,. 
\eea
This corresponds to the regime where the dipole is likely to have multiple interactions with the medium.

Eq.(\ref{eq:lam1}) will be equivalent to the Glauber –Gribov –Mueller (GGM) formula~\cite{Mueller:1989st}
\bea  \label{eq:lam1GGM}
\lambda_1 = \Gamma(\bar{y}_{\perp})\frac{ \alpha_s \pi^2}{N_C}b_{\perp}^2xG_N(x,1/b_{\perp})
 \,,
\eea
if we model the nucleus as a bunch of non-interacting nucleons with some effective Glauber gluon density $xG_N(x,1/b_{\perp})$ and a density  of nucleons per unit area $\Gamma(\bar y)$.  
Rescaling $k_{\perp} \to \hat k_\perp/ |b_{\perp}|$ in Eq.(\ref{eq:lammfp}) using 
Eq.(\ref{eq:MediumF}), we find an overall factor $\lambda_1\propto b_{\perp}^2$. 
We can then equate the $\hat k_\perp$ and $\bar{y}^-$ integral of the rescaled $\mathcal{B}$  with  $\Gamma(\bar{y}_{\perp})xG_N(x,1/b_{\perp})$; the number density of small $x$ gluons per unit area.  
$G_N$ absorbs an $\alpha_s N_c$, leaving $\alpha_s \pi^2/N_c$ as the prefactor.
This yields
Eq.(\ref{eq:lam1GGM}).
In the CGC formalism $xG_N(x,1/b_{\perp})$ is obtained by treating the nucleus as a source for small x gluons with radiative corrections from the BFKL/BK equations. The resummation modifies the saturation scale from its tree level value $Q_{s_0}$ to have a power law scaling with $x$; $Q_s(x) = Q_{s_0}x^{-2.44 \alpha_sN_C/\pi^2}$.  
The boundary condition for this resummation, in our case $\mathcal{B}(k_{\perp})$, is described in the CGC by the MV model \cite{McLerran:1993ni,McLerran:1993ka,McLerran:1994vd}.

\paragraph{{\bf Role of the medium size:}}

\begin{figure}
   \includegraphics[width=0.48\linewidth]{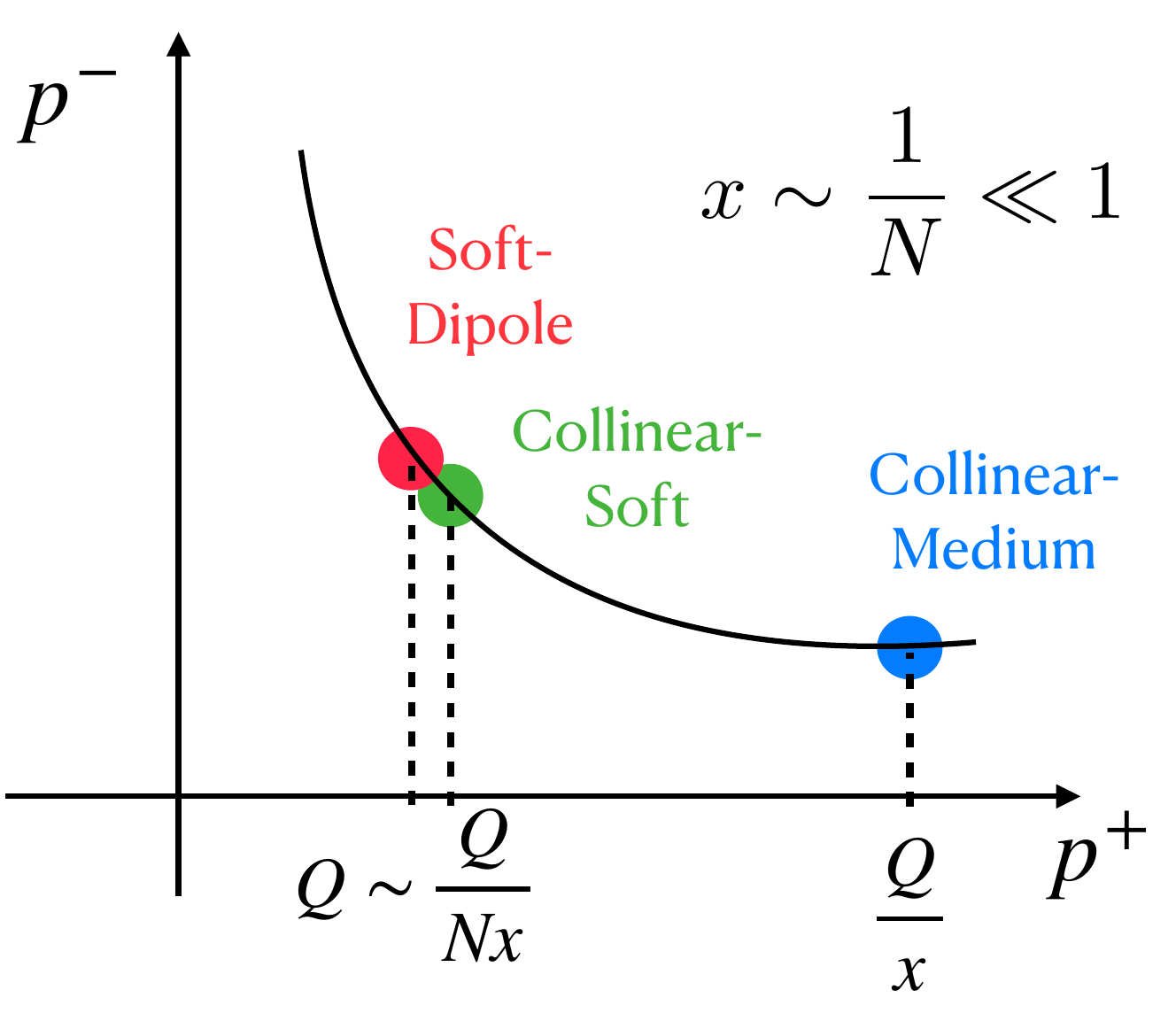}
\raisebox{-0.05cm}{   
   \includegraphics[width=0.48\linewidth]{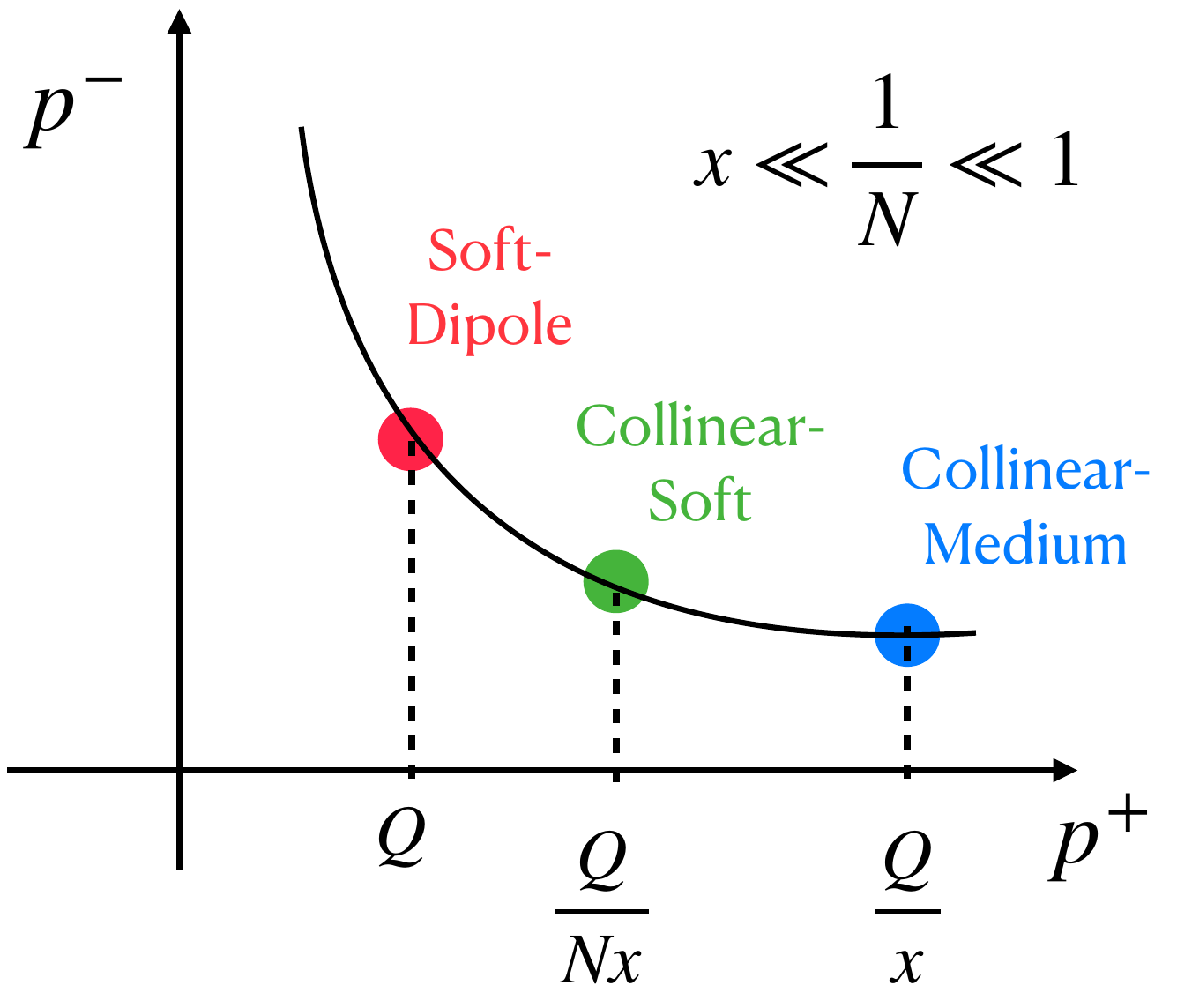}
}
   \includegraphics[width=0.54\linewidth]{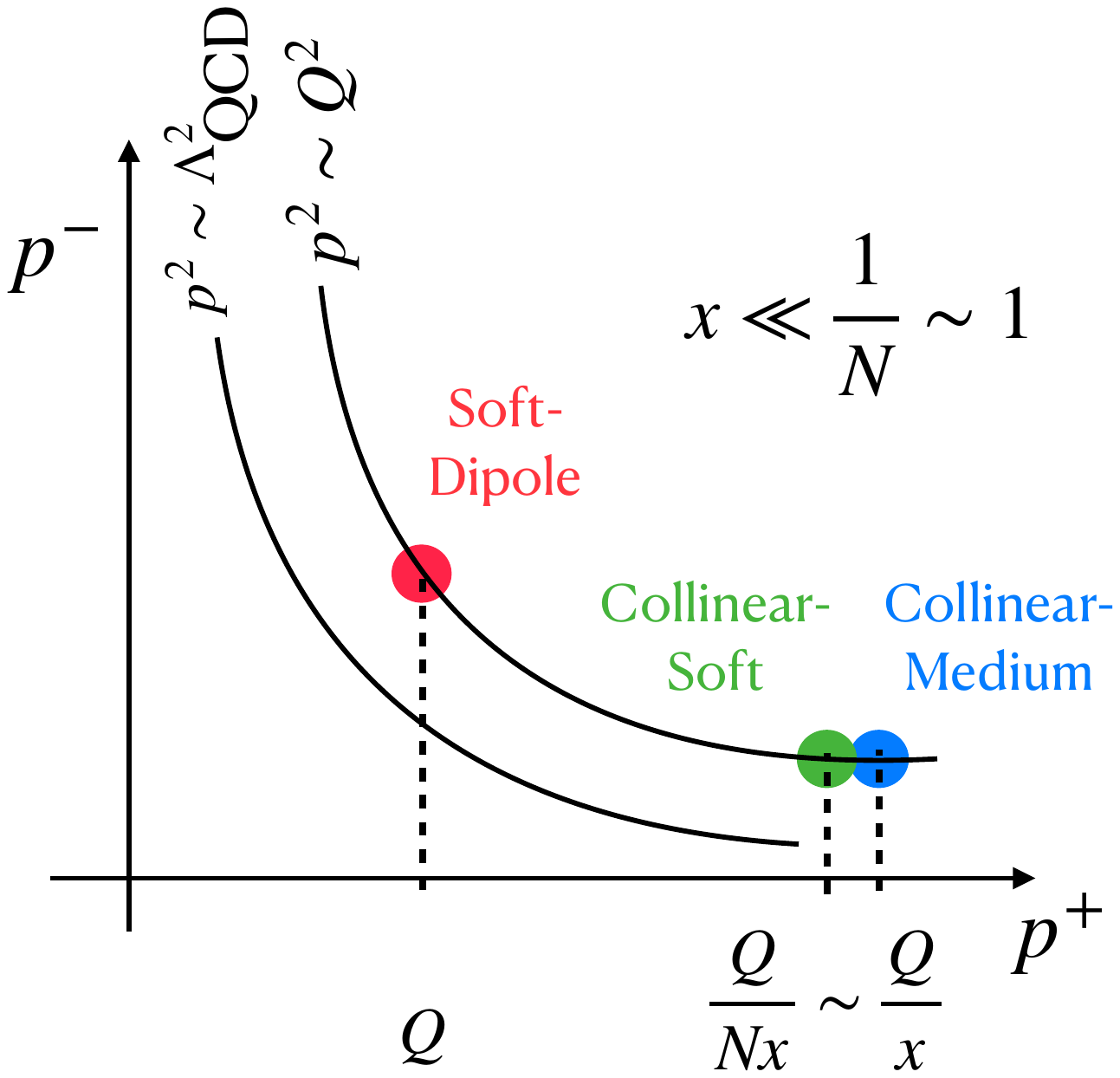}
\caption{Small $x$ regions for large $N$ nuclei (top panels) and small $N$ nuclei (bottom panel).}
  \label{fig:SC}
\end{figure}

From Eq.(\ref{eq:NLOG}),  we observe potential sensitivity to another 
length scale $L^-_N$; the extent of the medium in the $y^-$ direction. 
The relevant radiation sensitive to this scale will have momenta $p^+\sim  1/L^-_N$, and coherence time $t_c \sim L_N^-$. 
If $N$ is the typical number of nucleons seen by the probe along its path, then the rest frame size of the nucleus is $L_R\sim N/M_p$, and boosting yields $L_N^- \sim L_R  M_p Q/H \sim Q N/H$.  
Retaining a common virtuality $\sim Q^2$, this is a collinear-soft mode with
\bea
p_{cs} \sim \frac{H}{Q} \left( N x^2, \frac{1}{N} , x \right) .
\eea
We can consider two important regimes of the consequent EFT as we dial down the value of $x$ with fixed $Q$, keeping $N$ to be large but fixed as shown in Fig.~\ref{fig:SC}.    We see that at not so  small $ x\sim 1/N$,  
the soft and collinear-soft modes are the same so that the two mode factorization formula discussed above, Eq.(\ref{eq:NLOG}), is accurate.  

This means that while in a dense enough medium,  $\lambda_1 \gtrsim 1$ puts us in the saturation regime, it does not necessarily lead to a non-linear evolution.  If the medium is large ($L^-_N \sim 1/Q$)  we still have a linear BFKL evolution.

For smaller $ x\ll 1/N \ll 1$,  the collinear-soft and soft modes decouple, while still being well separated from the collinear function.  Even though we will leave the explicit derivation of the factorization formula in this regime for a future paper, we can still say a lot about the physics.
The dipole needs to interact at least twice with the nucleus to know its size, hence Eq.(\ref{eq:NLOG}) is modified for $m\geq$2. In particular, the soft function $S_m^{\mu\nu}$ must be further factorized into a $\bar y^-$(medium size) insensitive soft function $ \hat{S}_m^{\alpha\beta}$ (with $L_N^-\to 1/\nu$) and a collinear-soft function sensitive to $L_N^-\nu$. The $m$ copies of the function $\mathcal{B}_{\bar n}$ remains unchanged and still obey the BFKL equation with the same natural rapidity scale $\nu \sim Q/x = H/Q$.  The refactorization will modify the rapidity RG equation for the soft function away from BFKL with a natural rapidity scale $ \sim Q$.  However, by RG consistency the combination of collinear-soft and soft function must still obey the BFKL equation.  The natural rapidity scale for the CS function is $\nu\sim H/(QN)$, so evolution to the soft function sums logs of $Q^2N/H\sim xN$. Thus $\ln x$ summation will now have two contributions:  the linear BFKL piece from $\mathcal{B}_{\bar n}$; and that from the CS function. At leading order the latter modifies terms with at least two copies of $\mathcal{B}_{\bar n}$, so we expect it to reproduce the non-linear Balitsky–Kovchegov(BK) \cite{Balitsky:1995ub,Kovchegov:1999yj}/JIMWLK equation \cite{Jalilian-Marian:1997qno,Weigert:2000gi,Iancu:2000hn}.  
As a bonus, the  CS function's RG evolution also sums $\ln N$ terms. The linear to  non-linear evolution refers to the transition from pure BFKL to a nonlinear evolution which allows for both splitting and merging of what are referred to in literature as BFKL  pomeron which in our case is the BFKL resummed  $\mathcal{B}$ function.  The interference generated between successive interactions with the medium through a collinear soft mode, would allow this merging and splitting thereby leading to terms in the evolution equation with a higher powers of the $\mathcal{B}$ function , going beyond linear BFKL proportional to just a single $\mathcal{B}$. 

Finally,  as we approach $N=1$, the collinear soft mode merges with the collinear mode. 
The $O(m)$ glauber exchange yields a two mode factorization involving the same $\bar y^-$ insensitive Dipole function $\hat{S}_m^{\alpha\beta}$ mentioned in the previous paragraph.
However, in this case the independent scattering approximation breaks down, so rather than $m$ copies of ${\cal B}_{\bar n}$ two-point correlators, we now have $m$ point correlators ${\cal B}^m_{\bar n}$. This dependence on distinct  boundary conditions for each $m$ seems to complicate the resummation. 
To determine from our EFT whether the $m$ dependence of these boundary conditions is calculable at $Q$ with a small number of nonperturbative functions at $\Lambda_{\rm QCD}$ requires a further scale separation $\Lambda_{\rm QCD}^2\ll Q^2$ that we have not performed here.
This assumption has been implemented when the CGC formalism has been applied to a single hadron~\cite{Iancu:2015joa,Beuf:2020dxl}. 

We setup a top-down EFT for small $x$ DIS on a large inhomogeneous nucleus, with specific power counting parameters that define its range of validity. We gave a factorization formula which has an operator definition for the universal physics of the medium, and separates out the probe and small-$x$ evolution.  The structure at all orders in the number of independent probe-medium interactions leads us to define the mean free path of the probe by an operator matrix element in the nuclear state. This yields a power counting parameter that controls the importance of multiple interactions with the medium, and a means to define the saturation scale. The case $x\ll 1/N \ll 1$ requires the introduction of a new soft-collinear radiation mode, 
providing a distinct condition for understanding the transition to non-linear evolution in small-$x$ resummation.

\begin{acknowledgments}
We thank Ira Rothstein and Raju Venugopalan for helpful discussions.
This work was supported by the U.S.~Department of Energy, Office of Science, Office of Nuclear Physics, from DE-SC0011090 and in part by the Simons Foundation through the Investigator grant 327942. V.V is supported by startup funds from the University of South Dakota.
\end{acknowledgments}

\bibliography{bib}

\end{document}